\documentstyle[11pt,a4]{article}
\begin{document}
\vspace{4cm}

\title{The Growth of Interaction Cross-Section of 4.3 GeV 
Electrons in Excited Crystals}

\author{A.Aganyants}

\maketitle

\begin{center} 
{
Yerevan Physics Institute, Alikhanian 
Brothers Str.2, \\
Yerevan 375036, Armenia 
}
\end{center}

\abstract 
{
 The microscopic properties of the interactions of 
relativistic electrons were observed to alter when 
macro-conditions, such as the crystal vacuum, were changed. 
A qualitative explanation of the phenomenon is given. 
}

\newpage

\section*{Introduction}

The present paper is a continuation of earlier experimental works on the 
interaction of relativistic electrons with excited crystal vacuum as the 
physical medium. The first observations of the effect were carried out 
in 1976, delivered at a conference \cite{ag1} and reported in 
papers \cite{{ag2},{ag3},{ag4}}.

Briefly, the observed phenomenon consists essentially in the fact that 
beginning from some threshold intensity of the electron beam passing 
through a single crystal, the radiation increases with the beam intensity. 
The results of measurements point out that the emitted intense radiation 
is due to the interaction of bunch as a whole with the crystal. 
This radiation is produced on the boundary of crystal 
(like the transition radiation) that is excited by the same beam, 
and below we shall name that as the vacuum radiation. It is basically low 
energy radiation \cite{ag4} and proved to be by many times more intense 
than the channeling radiation under the same condition \cite{ag3}.

In the present paper we provide new more explicit experimental data in 
support of the electron beam intensity effect. The measurements 
have been taken on the internal electron beam of the Yerevan Synchrotron with 
the same experimental setup \cite{{ag4},{ag5}}. Below we give measurement
data obtained using two different techniques with better 
monitoring of electron and gamma-beam intensities:

\section{Experimental results}

\subsection{Technique one}

Data measured by means of ionization chambers, where the chamber was 
placed at the distance of 20 m from the diamond crystal downstream the 
beam-line, and the beam was collimated (by 3$\times$3 mm$^2$ aperture) 
at 10 m from the same crystal. 
The $\gamma-$ beam was cleaned of charged particles and passed parallel 
to the chamber electrodes (in the air gap of 1.5 {\it cm}). 
The electrons (positrons) that converted on air from gamma-quanta of energies
up to the edge of bremsstrahlung spectrum contributed to the ionization 
current. Despite high background from higher energy photons, the contribution 
of low energy photons from the oriented crystal proved to be predominant. The 
data obtained with the help of ionization chamber $N_{i}$ were normalized 
to those of quantameter $N_{q}$, which was gamma-beam intensity monitor. 
The dependence of ratio $N_{i}/ N_{q}$ on the angle ${\psi}$
of electron incidence on (110) planes is given in Table \ref{tab:tab1} 
Two rightmost columns correspond to the case, when 100${\mu}$m thick
aluminum converter was installed in front of the ionization chamber.  
Though the ratio $N_{i} / N_{q}$ grows both at the 
disoriented and oriented ${\psi}=$0 crystals when the electron 
intensity is increased, in the latter case the effect is more pronounced. 
It is worthwhile to note that though the effect appears to be small, 
in reality it is rather strong since, as was mentioned above, 
all photons of the bremsstrahlung spectrum, and not only the low energy ones, 
make contribution to the current of ionization chamber. The ionization 
chambers has the advantage that in contrast to the photomultiplier 
tube base detectors, it is an integral action instrument that like 
the quantameters may be operated at high particle fluxes.

\vskip 0.2cm

\begin{table}[ht]
\begin{center}
\caption{ }

\begin{tabular}{|l|c|c|c|c|}             \hline

Intensity, ${\gamma}-$quanta/sec & 1060 & 2500 & 1000 & 3000 \\ \hline
Chamber converter&\multicolumn{2}{c|} {Air}&
\multicolumn{2}{c|}{Aluminum foil+Air}  \\ \hline     
  
Disoriented crystal& 1.23 &1.29& 1.39 &1.69         \\ \hline

${\psi}=$ 0 &1.20 & 1.60& 2.11& 3.16       \\ \hline

\end{tabular}
\label{tab:tab1}
\end{center}
\end{table}

%\vskip 0.2cm

\subsection{Technique two}

The radiation losses of electrons were also detected by the missing 
energy of electrons. Here the synchrotron {\it per se} served as a
spectrometer 
of missing energy. In such a case the detection is made within a narrow 
energy range and the intensity effect is strongly manifested. 
An increase in instantaneous beam intensity  ({\it i.e.}, the beam
density) is 
achieved by inducing a faster beam dump. The electrons that lose some 
threshold energy in the crystal may be thrown out by the beam dump on the 
walls of the synchrotron vacuum chamber primarily near the target crystal. 
The produced showers will be partly detected with a scintillation 
counter placed near the crystal. The detector had a voltage lower than 
nominal to have a higher counting rate capability. The relative monitoring 
of electrons interacting with the crystal has been made with the quantameter. 
The measurement data are given in Tables \ref{tab:tab2},\ref{tab:tab3} for 
two orientations of the crystal. There is no need for giving 
the orientation dependence on the whole as was made in \cite{ag4}. 
A sharp increase in the counting rate with electron intensity ($\sim$ 
3 times) is seen in Table \ref{tab:tab2} for crystal orientation 
${\psi}=$ 0. A sharp increase is observed even at the orientation 
${\psi}=$ 0.46 mrad. 
The same is seen in Table \ref{tab:tab3} with data taken using another 
scintillation counter with the  gamma-converter placed in the 
experimental hall at the distance of 30 m from the diamond crystal, 
the gamma-beam being collimated by 6.7$\times$ 6.7 mm$^2$ aperture. 
As the energy acceptance of detector here is wider than in 
Table \ref{tab:tab2}, the integral effect is naturally less pronounced. 
The produced results explicitly confirm the presence of electron 
beam intensity effect showing up as an increase in the radiation 
cross-section of 4.3 GeV in diamond single crystal.
 
\begin{table}[ht]
\begin{center}
\caption{}

\begin{tabular}{|l|c|c|}   \hline

Beam dump & Slow dump &  Faster dump      \\ \hline
Disoriented crystal& 284 $\pm$ 9&6.7$\cdot$10$^3$   \\ \hline
Oriented crystal, ${\psi}=$ 0.46 mrad.&1.2$\cdot$10$^3$& 
6.6$\cdot$10$^5$        \\ \hline
Orientation ${\psi}=$ 0& 5.2$\cdot$10$^3$&10$^6$  \\  \hline   

\end{tabular} 
\label{tab:tab2}
\end{center}   
\end{table}

\begin{table}[ht]
\begin{center}
\caption{}

\begin{tabular}{|l|c|c|}   \hline

Beam dump& Slow dump&  Faster dump       \\ \hline
Disoriented crystal& (0.58 $\pm$ 0.02) $\cdot 10^3$& 1.7$\cdot 10^3$   
\\ \hline
Oriented crystal, ${\psi}=$ 0.34 mrad.&1.8$\cdot$10$^3$ &
5.26$\cdot$10$^3$   
\\ \hline
Orientation ${\psi}=$ 0& 3.9$\cdot$10$^3$&17.5$\cdot$10$^3$  \\  \hline

\end{tabular}
\label{tab:tab3}
\end{center}
\end{table}

\section{Conclusion}
                                               	
One can interpret this phenomenon as a result of electron interactions 
with a nonlinear medium such as an excited crystal. It may be 
qualitatively explained as follows: the first relativistic electrons 
of a bunch travelling through a crystal knock out electrons of atomic 
shells, as a result of which the electromagnetic vacuum near these atoms 
is excited. In case of high intensity electron beam the correlations 
of excitations in this non-equilibrium process \cite{pri6} are
strengthened 
and grow macroscopic in the range of characteristic frequencies 
${\omega}_{ex}$. This may mean a spontaneous symmetry breaking of 
the vacuum \cite{lee7}, 
in this case of electromagnetic vacuum. As a result, the permittivity 
of medium may be $\varepsilon < 1$ in the range  ${\omega}_{ex} < 
{\omega}<{{\omega}_{ex}}{\gamma}$
(where ${\gamma}$  is the Lorentz-factor of relativistic electrons). 
When the other relativistic electrons of the bunch enter the medium 
with changed structure of that kind, they will emit photons in the 
mentioned frequency range due to virtually polarized atoms similar to 
the transition radiation \cite{jac8}, the frequencies of which are 
considerably lower.
For example, the energy of $K_{\alpha}$ transition in carbon (the
diamond) is 283 eV,
 but that of the plasma frequency is $\sim$20 eV. The frequencies  
${\omega}_{ex}$ and ${\omega}$
for substances with larger atomic numbers $Z$  must be still higher. 
Therewithal the intensity of radiation will be determined by the degree 
of vacuum excitation. The faster-than-quadratic increase of
radiation 
cross-section in Table 2 is, apparently, due to the non-equilibrium 
phase transition in the excited vacuum.
The periodicity of crystal structure and its orientation influence in such 
conditions both the radiation spectra and angular distributions 
(anomalous wide one \cite{ag9}) of gamma-quanta, as the density of
substance is 
increased in the directions of crystal planes and axes. In our experiments 
the intense vacuum radiation has been really observed in the range of several 
MeV \cite{ag4}.  Although the radiation yields decrease at higher energy
of photons, 
the influence of the medium excitation is appreciable up to the end of the 
bremsstrahlung spectrum \cite{ag9}. All these facts do not contradict the 
aforementioned interpretation. The measurements of radiation spectra 
in terms of electron beam intensity effect and the theoretical treatment 
of radiation mechanism are, however, of current interest. 
It is noteworthy that the subject of this work transgresses the bounds 
of purely radiative processes, as they happen in strongly non-equilibrium 
medium, that takes place, for example, in relativistic heavy ion collisions 
(the citing of Refs \cite{{pri6},{lee7}} here was not accidental). 
In this sense the 
subjects touched upon in the paper are of interest for many fields of physics.
Not to go into details, we should only like to mention that this vacuum 
radiation was used since 1976 for finding the planes and axes of crystals. 
It is obvious that this vacuum radiation may find wide application:

\begin{enumerate}

\item As a source of regenerated positrons in future  electron-positron 
linear colliders;

\item As a source of monochromatic gamma-quanta in the range 
of 5 - 50 keV in small electron accelerators with energies 20 - 100 MeV.

\item For preparation of an active medium for potential gamma-lasers.

\end{enumerate}

\section*{Acknowledgments}
In conclusion the author thanks Avakian R., Berman B. and 
Ter-Michaelyan M. for discussions.

%\vskip 0.2cm

\newpage

\section*{Table Captions}
\begin{description}
\item[Table 1.] Normalized integrated ionization currents $N_{i}/N_{q}$   
versus  the electron beam intensity with two different
converters of gamma-quanta at two different
orientations of 72 ${\mu}$m thick diamond crystals.
The statistical errors were neglected.

\item[Table 2.] Counting rates of the detector for some orientations of
crystal for different intensities of electron beam. The diamond crystal 
was 100 ${\mu}$m thick.

\item[Table 3.] Counting rates of the detector placed in the
experimental
hall for some orientations of 100 ${\mu}$m thick
crystal for different intensities of electron beam.
\end{description}

\end{document}